\documentclass[fleqn,10pt,twoside]{article}

\usepackage[english] {babel}

\usepackage {graphicx}
\usepackage {graphics}
\usepackage {floatflt}
\usepackage {latexsym}
\usepackage {caption3} %Пакет для MikTex 2.6, где содержатся команда для замены двоеточия на точку в подписи к рис., например, Рис.1: -> Рис.1.
\usepackage[labelsep=period]{caption}[2007/11/04 v.3.1e]

\def\noi{\noindent}

\renewcommand{\thesubsubsection}%
        {\arabic{section}.\arabic{subsection}.\arabic{subsubsection}.}

\newcommand{\heads}[2]{\markboth{\protect\small\it #1}{\protect\small\it #2}}

\newcommand{\Arthead}[5]{ \setcounter{page}{#4}\thispagestyle{empty}\noi
    \unitlength=1pt \begin{picture}(500,40)

        \put(0,58){\shortstack[l]{\small\it Gravitation \& Cosmology,
                        \small\rm Vol. #1 (#2), No. #3, pp. #4--#5    \\
        \footnotesize {Proceedings of the International Conference on Gravitation, Cosmology, Astrophysics and Nonstationary Gas Dynamics,}    \\
\footnotesize {Delicated to Prof. K.P.Staniukovich's 90th birthday, Moscow, 2-6 March 2006}    \\
\footnotesize\copyright \ #2 \ Russian Gravitational Society} }

    \end{picture}
	 }     		%%% #1: volume; #2: year; #3: issue; #4-#5: pages
\def\prepno#1#2
    {\thispagestyle{empty}
    \noi    \unitlength=1mm
    	\begin{picture}(178,10)
            \put(177,15){\llap{\bf #1}}
            \put(177,10){\llap{\small\rm #2}}
        \end{picture}
          }             %%% #1: prepr.no.; #2: journal-ref

\newcommand{\Title}[1]{\noi {\uppercase{\Large #1}}     }%\\}
\newcommand{\Author}[2]{\noi{\large\bf #1}\\[2ex]\noindent{\it #2}   }%\\}

\newcommand{\Abstract}[1]{\vskip 2mm \begin{center}
        \parbox{16.4cm}{\small\noi #1} \end{center}\medskip}

\newcommand{\foom}[1]{\protect\footnotemark[#1]}

\newcommand{\email}[2]{\footnotetext[#1]{e-mail: #2}
		\addtocounter{footnote}{1}}

\heads{A.M.Baranov and R.V.Bikmurzin}
      {Exact static solutions for fluid gravitating balls  in homogeneous coordinates}

\begin{document}
\twocolumn 
[
\Arthead{12}{2006}{2-3 ({\bf{46-47}})}{103}{105}

\Title{EXACT STATIC SOLUTIONS FOR FLUID GRAVITATING BALLS %\\

\vspace{0.2cm}
IN HOMOGENEOUS COORDINATES}

\vspace{.5cm}
   \Author{A.M.Baranov\foom 1 and %%
  R.V.Bikmurzin\foom 2}   %%   If there is
{\it Dep. of Theoretical Physics,Krasnoyarsk State University,
79 Svobodny Av., Krasnoyarsk, 660041, Russia}

{\it Received 22 April 2006}

\Abstract
{Two new classes of exact interior static solutions of the Einstein equations 
in homogeneous coordinates for a gravitating  ball filled by a Pascal perfect fluid are obtained. Schwarzschild's interior solution of is a special case of these solutions.}
]

\email 1 {bam@lan.kras.ru}
\email 2 {Rustam\_B\_V@mail.ru}

\section{Introduction}

The choice of a coordinates' system can simplify the representation of the Einstein equations. A spherically symmetric metric interval in 4D space-time is usually written in of the curvature coordinates. In this paper, we make use of the homogeneous coordinates according to the Weyl remark [1] that the 3D spherical metric interval is conformally flat in such coordinates. So the static spherically symmetric metric interval for an interior gravitational field of a fluid stellar model with a general distribution of mass density is written as 
$$
ds^2=A(r) dt^2-B(r) \left( dr^2+r^2(d\theta^2+\sin^2 \theta \,d\varphi^2)\right)=
$$
$$
=R_0^2(A(x)d{\tau{^2}}-B(x)dl^2).
\eqno{(1)}
$$

\noindent
Here we introduce the dimensionless coordinates $\tau = t/R_0;\,x=r/R_0;\, 
0 \leq x \leq 1,$ where $t$ is the time variable; $r$ is the radial variable; $\theta; \varphi$ are angular variables; $R_0$ is a stellar exterior radius; $dl^2=
(dx^2+x^2(d\theta^2+\sin^2 \theta \,d\varphi^2));$ the velocity of light $c=1;\,$ $A(x)$ and $B(x)$ are metric functions.

\section{The first class of solutions}

The Einstein equations for the energy-momentum tensor of Pascal's perfect  fluid 
$$
T_{\alpha\beta}=\left(\mu (x)+p(x) \right)U_{\alpha}U_{\beta}-p(x)\,g_{\alpha\beta} \eqno{(2)}
$$

\noindent
are transformed to three equations for $A(x)$ and $B(x)$ functions.
The 4-velocity in a comoving frame of reference is equal to $U_\alpha=\delta^0_{\alpha} \sqrt{A(x) R_0^2}.$

In the case with $\mu=const\,$ (a homogeneous distribution of mass),
one of these equations may be rewritten as Emden's equation [2]
$$
\frac{d^2 b(x)}{dx^2}+\frac{2}{x} \frac{d b(x)}{dx} + b(x)^5=0.
\eqno{(3)}
$$ 

From this equation we can find the function $B(x)$ and then, from other gravitational equations, we obtain function $A(x)$ [3]
$$
B_{Sch.}(x)=\frac{C_1}{(C_2+x^2)^2}, 
$$
$$
A_{Sch.}(x)=\left[C_4- \frac {C_3} {(C_2+x^2)} \right]^2.
\eqno{(4)}
$$

Thus we have obtained Schwarzschild's interior solution in homogeneous coordinates. The Weyl tensor is equal to zero for such exact solution which describes the conformally flat 4D space-time. The junction conditions on the stellar surface give connections of constants with compactness $\eta=2m/R_0,$ where $m$ is Schwarzschild's mass and $0 \leq \eta < 1$.

When $\mu(x)$ is a function of $x,$ we take the function $B(x)$ as a generalization of the case with a homogeneous distribution of mass density in the form 
$$
B_{in}(x)=\frac{C_1}{(C_2+x^2)^k}=C_1 Y(x)^{-k},
\eqno{(5)}
$$

\noindent
where $Y(x)=\left(C_2+x^2\right)$ and $0 \leq k \leq 2$, then we can get functions $A_{in}(x)$, $\mu(x)$ and $\:p(x)$:
$$
A_{in}(x)=\left[ \frac {C_4Y(x)^{\Lambda}-C_3}{Y(x)^{(k+\Lambda-1)/2}}\right]^2; \eqno{(6)}
$$
$$
\mu(x)=\frac {k Y(x)^{k-2}}{8\pi R^2_0 C_1} \bigl[(2-k)Y(x)+(k+4)C_2\bigr]; \eqno{(7)}
$$
$$
p(x)=\frac{Y(x)^{k-2}}{4\pi R_0^2 C_1} \biggl[ Y(x) + 7k C_2+ \frac{C_3+C_4Y(x)^\Lambda}{C_3-C_4Y(x)^\Lambda}\times 
$$
$$
\times\Lambda\bigl((k-1)Y(x)-kC_2\bigr) \biggr] -3\mu (x),
\eqno{(8)}
$$

\noindent
where $\Lambda=\sqrt{2k^2-4k+1}\geq 0$ and the parameter $k$ can have values 
$ k=\left(0;1-1/\sqrt{2}\right] \cup \left[ 1+1/\sqrt{2};2\right].$

In this case, the functions $B(x)$ and $\mu(x)$ are chosen as nonsingular functions for the stellar source. We  also take the pressure equal to zero on the stellar surface.

So the metric interval (1) may be rewritten as 
$$
ds^2=\frac{R^2_0 C_1}{Y(x)^{k}}  \left[ \frac{Y(x)^{1-\Lambda}}{C_1} (C_4 Y(x)^\Lambda-C_3)^2 d\tau^2 - dl^2 \right]
\eqno{(9)}
$$

Now we must find the constants $C_1$, $C_2$, $C_3$ and $C_4$ for a final result.

As an exterior solution, we take the Schwarzschild exterior solution in homogeneous dimensionless coordinates (see, e.g., [4])
$$
A_{ext}(x)=\left( \frac{4x-\eta}{4x+\eta}\right)^2, \quad
B_{ext}(x)=\left( 1+\frac{\eta}{4x}\right)^4.
\eqno{(10)}
$$

Using the smooth junction conditions on the 2D stellar surface with the exterior solution in the form (10) in homogeneous coordinates, we have 
$$
C_1=\left( 1+\frac{\eta}{4} \right)^4 \left[ \frac{k(4+\eta)}{2\eta} \right]^k; \eqno{(11)}
$$
$$
C_2=\frac{k(4+\eta)}{2\eta} -1;
\eqno{(12)}
$$
$$
C_3=\frac{k}{4\eta \Lambda} \Bigl( 4k-(\eta-4)(k-1-\Lambda) \Bigr) \left[ \frac{k(4+\eta)}{2\eta} \right]^{(k-3+\Lambda)/2};
\eqno{(13)}
$$
$$
C_4=\frac{k}{4\eta \Lambda} \Bigl( 4k-(\eta-4)(k-1+\Lambda) \Bigr) \left[ \frac{k(4+\eta)}{2\eta} \right]^{(k-3-\Lambda)/2}.
\eqno{(14)}
$$

From a requirement $C_2 > 0$ it follows 
$$
\frac{k(\eta+4)}{2\eta}-1>0 \;\; \Rightarrow \;\; k>\frac{2\eta}{\eta+4} \;\; \mbox{or} \;\; \eta<\frac{4k}{2-k}.
\eqno{(15)}
$$

If the parameter $k$ is in the range $\left(0;1-1/\sqrt{2}\right]$, then the 
compactness $\eta$ has values in the interval \linebreak
$\left[0;\;4(\sqrt{2}-1)/(\sqrt{2}+1) \right) \approx \left[0;\;0.686 \right).\,$ If $k>1+1/\sqrt{2},$ then parameter $\eta$ is arbitrary.

\begin{figure}[t,h]
%\begin{center}
\fbox{\parbox{8.2cm}{\rule[-0.5cm]{0 mm}%
{8.2cm}\hfil\centering\includegraphics[width=8cm]{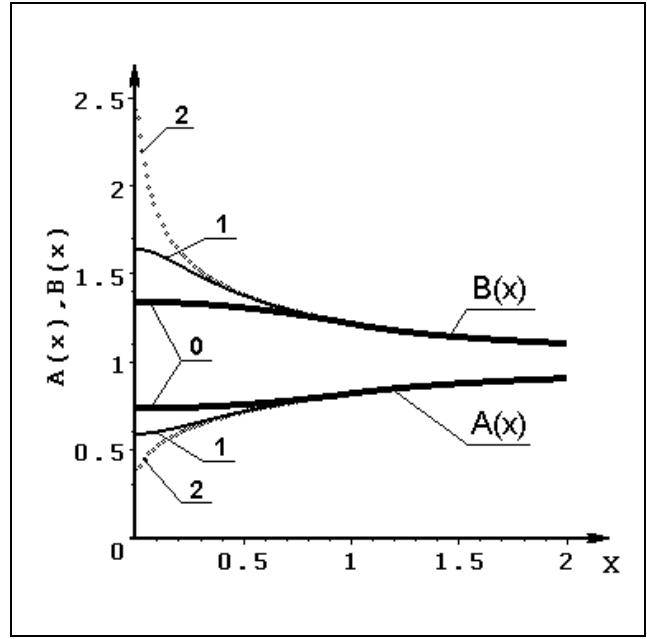}}}
\caption{Behavior of the metric functions $A(x)=g_{00}(x)$ and $B(x)=g_{11}(x)$ of the complete solutions belonging to the first and second classes with $\eta=0.2$ and $k=0.1003$. The number 0 corresponds to Schwarzschild's interior solution. The numbers 1 and 2 correspond to the first and second classes of solutions.}
%\end{center}
%\label{fig:fig1}
\end{figure}

As a result, all functions of the first-class solutions may be rewritten 
in a clear form: 
$$
B_{in}(x)=\left(1+\frac{\eta}{4}\right)^4 \left[ 1-\frac{2\eta}{k} \frac{(1-x^2)}{\eta+4}\right]^{-k}=
$$
$$
=\left( 1+\frac{\eta}{4} \right)^4 \Psi (x)^{-k};
\eqno {(16)}
$$
$$
A_{in}(x)=\biggl[ \frac{(\eta-4)(k-1-\Lambda)-4k}{2\Lambda(\eta+4)}\Psi(x)^{(1-k\Lambda)/2} \times
$$
$$
\times \bigl( 1-\xi \Psi(x)^\Lambda \bigr) \Bigr]^2;
\eqno{(17)}
$$
$$
\mu(x)=\frac{128\eta}{\pi R_0^2k}\frac{\Psi(x)^{k-2}}{(\eta+4)^6}\bigl[ 12k+\eta(k-2)(3-x^2) \bigr],
\eqno{(18)}
$$

\noindent
where the following notations are used: 
$$
\xi=\frac{(\eta-4)(k-1+\Lambda)-4k}{(\eta-4)(k-1-\Lambda)-4k};
$$
$$
\Psi(x)=\biggl[ 1-\frac{2\eta}{k}\frac{(1-x^2)}{(\eta+4)}\biggr].
\eqno{(19)}
$$

\section{The second class of solutions}

Now we shall generalize the function $A(x)$ from (4) by a similar way.
The function $A(x)$ is defined as
$$
A_{in}(x)=C_1(C_2-x^2)=C_1Y(x)^k.
\eqno{(20)}
$$

Then the Einstein equations give
$$
B_{in}(x)=\left[ \frac {Y(x)^{(\Lambda-1-k)/2}}{C_4Y(x)^{\Lambda}-C_3}\right]^2; 
\eqno{(21)}
$$
$$
p\:(x)=\frac {Y(x)^{k-\Lambda-1}}{8\pi R_0^2}   
\biggl[ \Bigl( k^2\bigl(Y(x)-C_2\bigr)-2C_2(1+\Lambda)\Bigr)\times
$$
$$
\times\bigl(C_3+C_4Y(x)^\Lambda\bigr)^2
 + \Bigl( Y(x)(1+k^2)+C_2(1+\Lambda-k^2)\Bigr)\times
$$
$$
\times 4C_3C_4Y(x)^\Lambda +4C_2C_3^2\Lambda \biggr];
\eqno{(22)}
$$

$$
\mu(x)=\frac{k\bigl(C_3-C_4Y(x)^\Lambda \bigr)}{4\pi R_0^2Y(x)^{1+\Lambda-k}} \biggl[ \Bigl(3C_2-\Lambda\bigl(Y(x)-C_2\bigr)\Bigr) \times
$$
$$
\times\bigl(C_3-C_4Y(x)^\Lambda\bigr) + 2\Lambda C_3(Y(x)-C_2) \biggr]-3p(x),
\eqno{(23)}
$$

\begin{figure}[t,h]
%\begin{center}
\fbox{\parbox{8.2cm}{\rule[-0.5cm]{0 mm}%
{6.5cm}\hfil\centering\includegraphics[width=8cm]{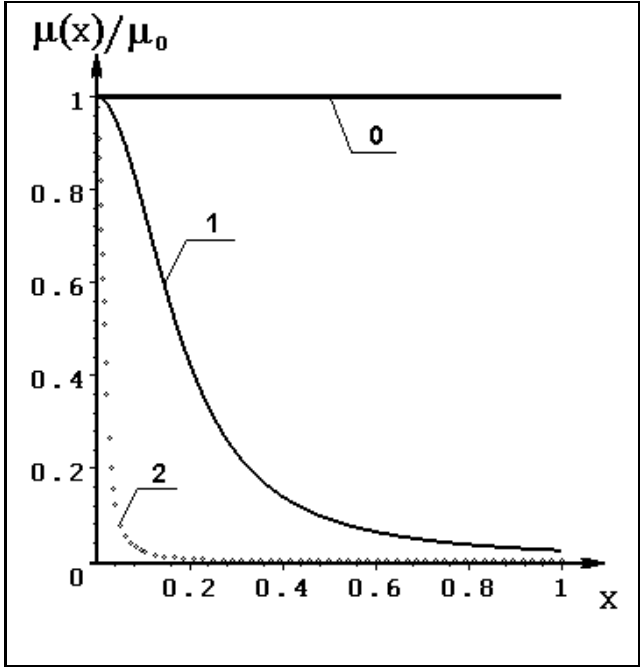}}}
\caption{Behavior of the mass density $\mu(x)$ of the total solutions belonging to the first and the second classes with $\eta=0.2,$ $k=0.1003$ and a central mass density $\mu_0=const.$ The number 0 corresponds to Schwarzschils's interior solution. The numbers 1 and 2 correspond to the first and the second classes of solutions.}
%\end{center}
%\label{fig:fig1}
\end{figure}

\noindent
where $\Lambda=\sqrt{1+2k^2}.$

Under the smooth junction conditions with Schwarzschil's exterior solution, the new constants are
$$
C_1=\left(\frac{4-\eta}{4+\eta}\right)^2\left[ \frac{8\eta}{k(16-\eta^2)}\right]^k;
\eqno{(24)}
$$
$$
C_2=\frac{k(16-\eta^2)}{8\eta}-1;
\eqno{(25)}
$$
$$
C_3=\frac{2\bigl(4(\Lambda+1)+k\eta)}{(\eta+4)^2\Lambda}\left[ \frac{8\eta}{k(16-\eta^2)} \right]^{(1+k-\Lambda)/2};
\eqno{(26)}
$$
$$
C_4=\frac{2(4(\Lambda-1)-k\eta)}{(\eta+4)^2\Lambda }\left[ \frac{8\eta}{k(16-\eta^2)} \right]^{(1+k+\Lambda)/2}.
\eqno{(27)}
$$

Finally, we have bounds for the parameter $k$
$$
\frac{80+40\eta-7\eta^2+\chi}{\eta^3-10\eta^2-64\eta+480} 
\geq \frac{k}{2} \geq \frac{4\eta}{(16-\eta^2)},
\eqno{(28)}
$$

\noindent
where $\chi$ is
$$
\chi=\sqrt{17\eta^4-240\eta^3+2528\eta^2-8960\eta+6400}.
\eqno{(29)}
$$

\section{Summary}

The two classes of interior exact solutions of Einstein's equations, found in the paper, describe the gravitational field of a ball filled with a Pascal perfect fluid. These solutions correspond to a static stellar model which is a generalization of Schwarzschild's interior solution. Limiting conditions for the parameters have been found. The results of this study have been applied to a neutron star model with $\eta=0.2$ and the parameter $k=0.1003.$ Plots of the metric functions and the mass density have been constructed.

\end{document}